\documentstyle[12pt,a4]{article}
\begin{document}
   
\title
{Galactic Synchrotron Foreground and the CMB Polarization Measurements}

\author{M.V.Sazhin, \\
Sternberg Astronomical Institute, Universitetsky pr.13,\\ 
  119899, Moscow, Russia, \\
  email: sazhin@sai.msu.ru
}
  
\maketitle

\begin{center}
{\bf Abstract}
\end{center}
The polarization of the CMBR represents a powerful test for modern
cosmology. It allows to break the degeneracy of fundamental cosmological
parameters, and also to observe the contribution of gravitational waves
background to the CMBR anisotropy. To observe the CMBR polarization
several experiments are either in progress or planned
and SPOrt is one of the most promising
planned by ESA \cite{cor99}. At the same time the observation of
the CMBR polarization is a difficult task and  one of the reasons is the
presence of polarized foreground emission. For instance, galactic
polarized synchrotron emission (according to some estimates) can
completely mimic the polarization of the CMBR. Nevertheless, one can use
mathematical properties of the spherical harmonics of the distribution of
radiation over the sky to separate different contributions. In this paper 
the mathematical properties of the polarized synchrotron foreground and   
the physical mechanism that produces it are discussed. The separation of
synchrotron polarization from the polarization generated by density
cosmological perturbations is discussed as well.

\vskip1cm

\section{Introduction}

Next year will be a decade since the first detection
of the CMBR anisotropy \cite{str92}, \cite{smo92}. Afterwards the CMBR
anisotropy  observed a wide angular range including the intermediate
angular  scales. The first and second Doppler peaks and its angular
spectrum were observed \cite{boo00}, \cite{boo01a}, \cite{boo01b}. It
allows to determine the spectrum of primordial cosmological perturbations,
proves the inflation theory and allows to extend our knowledge to the very
early Universe. At the same time  new problems arose requesting new
approaches. The most natural way to solve them is the measurement of the
CMBR polarization.

The polarization of the CMBR was generated during the recombination epoch
of our Universe at $z \approx 1000$. M.Rees recognized first that
polarization promized new tool of cosmological investigation. Afterwards
it was considered by many authors (see, for instance \cite{saz95},
\cite{ng96}, \cite{mel97}).

The polarization of the CMBR indeed can provide very important information
about
the early Universe either proving or disproving gravitational waves 
background originated in
the very early stages of our Universe. Moreover it can  resolve
the degeneracy of fundamental
cosmological parameters: density of matter, density of dark energy etc.   
At the same time the observation of the CMBR polarization is a very
difficult aim because it is expected to be at least 10 times
less the CMBR anisotropy and it is mimicked by the inhomogeneous
foreground polarization.

This contribution will be devoted to the
discussion of both the inhomogeneous foreground polarization and
the CMBR polarization generated by density cosmological perturbations.
The problem of separation of these component is discussed too.

The microwave foreground radiation of our Galaxy consists of three main
components:
\begin{itemize}
\item the synchrotron radiation (strong polarized component),
\item the free-free radiation (polarization is assumed to be negligible),
\item dust radiation  (the situation is uncertain, but presumably it is 
negligible \cite{set98}).
\end{itemize}

As far as the second component is unpolarized and the polarization of dust
emission is uncertain, these preliminary considerations is
devoted the contribution of galactic synchrotron to the CMBR Stokes
parameters, comparising the Stokes parameters of inhomogeneous
synchrotoron radiation with the Stokes parameters generated by
cosmological density perturbations.

\section{Synchrotron Radiation and its Polarization}

The angular distribution of the synchrotron  radioemission polarized   
component is very inhomogeneous (see, for instance, \cite{rei01}). It
can affect our ability to observe the polarization of the CMBR.

This section discusses main principles of the synchrotron radiation,
resulting from the circular motion of an electron around magnetic lines
(see, for instance \cite{gin64}, \cite{gin69}, \cite{wes59}).

The galactic synchrotron radiation in fact is produced by relativistic
electrons which move into interstellar magnetic field. As far as they are
relativistic, their energy is much higher than the rest energy of electron
$E >> m c^2$.

The charged particles in magnetic field move along a helical path centered
on magnetic line. The frequency of the circular motion
is determined by the magnetic field $H$ value. In case of relativistic 
particle the frequency is
determined by the ratio of magnetic field over the energy of a particle:
$$
\omega_H =\frac{e H}{mc} \frac{m c^2}{E}
$$

This is the frequency of circular motion of an electron, but the 
frequency of the radioemission is much higher. The maximum in the spectrum
of one emitting electron corresponds to the frequency
\begin{equation}
\omega_m = \frac{e H_p}{mc} (\frac{E}{m c^2})^2
\label{fm}
\end{equation}

\noindent
Here $H_p$ is the component of the magnetic field perpendicular to the
direction of the  velocity vector. The additional multiplication factor ${
\displaystyle E \over \displaystyle m c^2}$ is a consequence of the    
Doppler shift. Therefore, the maximal frequency is much higher than the
circulating frequency of the electron.
  
The galactic magnetic field determines both the intensity and the
frequency of the radiation. Distribution of magnetic field is  random and
inhomogeneous in our Galaxy. Therefore, both the intensity
and the frequency of the synchrotron radiation follow the angular
structure of distribution of the magnetic field \cite{wie01}.
The polarization of the synchrotron radiation results to be anisotropic
both in intensity and angular distribution.

First of all I would like to discuss the synchrotron radiation of an
electron and its Stokes parameters. One can plot perpendicular axes in an
observer's plan perpendicular to the direction of the
propagation of an electromagnetic wave. I designate them as ${\vec l}$ and
${\vec r}$. The intensity of any harmonic vibration of electromagnetic
field can be projected into these vectors. The intensity along ${\vec l}$
will be designated as $I_l$ and along ${\vec r}$ will be designated as 
$I_r$. The component describing the correlation of intensity
between $\vec l$ and $\vec r$ will be designated as $I_u$. The Stokes
parameters are: $I=I_l + I_r$ , $Q=I_l -I_r$, and $U= I_u$.
One additional Stokes parameter $V$ defines the cicular
polarization. The parameters $I_l$, $I_r$, and $I_u$ are used in the
theory of polarization of the CMBR ($V=0$).
The Stokes parameters,  $I_l$ and $I_r$ are used also in
the theory of synchrotron
radiation,  but
instead of $I_u$ the angle $\chi$ is more convenient in the synchrotron
theory and it defined as
$$
\tan 2 \chi = \frac{U}{Q}
$$

\noindent
in the interval $0 < \chi < \pi$. It is angle between
the vector $ \vec l$ and the main axis of the polarization ellipse.

The Stokes parameters of the synchrotron radiation of a particle can be
written as a function of the amplitude of magnetic field, of the angle 
between the magnetic field direction and the line of sight, and
of the dimensionless frequency of the radiation ${\displaystyle \nu \over
\displaystyle \nu_c}$. $\nu_c = 1.5 \nu_m$  is determined by (\ref{fm})
consequently depends of the magnetic field too  \cite{gin64}.

The electron moving around magnetic lines produces all Stokes parameters:
$I$ (intensity), $Q$ and $U$ (describing the linear polarization) ,
and $V$ (describing the circular polarization). However, relativistic
electron produce mainly linear polarization. More exactly the
degree of circular polarization over linear polarization is of the order
of
$$
\sim o(\frac{m c^2}{E}).
$$

The projection of magnetic vectors on the observer's plan defines the  
polarization ellipse, being  the minor axis of the ellipse  along the  
projection. The angle between the vector $\vec l$ and the major axis of
the polarization ellipse on the observer's plan is  $\chi$. The
distribution
of the angle $\chi$ is random inside interval $0 < \chi < \pi$ and
it is independent on the direction of observation.

The observed synchrotron radioemission is produced by an ensemble of 
relativistic electrons and as far as Stokes parameters are additive the
contribution of separate particles are summed. Although
the intensity of the radiation produced by an ensemble of particles having
distribution $N(E, \vec R, \vec k)$ can be found in many books devoted to
the subject, their Stokes parameters $Q$ and $U$  less usual and  they can
be rewritten as:
\begin{equation}
Q=w_0 \int dE dl N(E,l,\theta, \varphi, \vec k) H(l, \theta, \varphi) \sin
\mu(\theta, \varphi) \cos 2 \chi \frac{\nu}{\nu_c} K_{2/3}\left(
\frac{\nu}{\nu_c}\right)
\label{q1}
\end{equation}

\begin{equation}
U=w_0 \int dE dl N(E,l,\theta, \varphi, \vec k) H(l, \theta, \varphi) \sin
\mu(\theta, \varphi) \sin 2 \chi \frac{\nu}{\nu_c} K_{2/3}\left(
\frac{\nu}{\nu_c}\right)
\label{u1}
\end{equation}

\noindent
Here $w_0 = \frac{\displaystyle \sqrt{3}e^2}{\displaystyle m c^2}$ is a
constant, $H(\vec r) = H(l, \theta, \varphi)$ is spatial
distribution of the magnetic field,  $\mu$ is angle between the line of
sight and the magnetic vector ($H_p= H \sin \mu$), $ K_{2/3}$
is modified Bessel  function and $\nu_c$ was defined above.

The equation (\ref{q1}) differs from equation
(\ref{u1}) because the term $\cos 2 \chi$  is substituted by $\sin 2  
\chi$. This angle is random over the sky. 

Under reasonable assumptions one can get the Stokes parameters as
functions of frequency, magnetic field and angle $\chi$:

$$
Q= c_1 \left( H(l, \theta, \varphi) \sin
\mu(\theta, \varphi)\right)^{\gamma +1} \nu^{-\gamma} \cos 2 \chi
$$

\noindent

$$
U=c_1 \left( H(l, \theta, \varphi) \sin
\mu(\theta, \varphi)\right)^{\gamma +1} \nu^{-\gamma} \sin 2 \chi
$$

\noindent
Here $\gamma$ and $c_1$ are constants related with the energy law of
relativistic particles distribution etc.

What is very important for our considerations is that   
both $Q$ and $U$ are assosiated to the synchrotron radiation of our Galaxy
and they can be rewritten as a function $F(\theta,
\varphi)$ of the magnetic field, distribution of relativistic particles
etc, which depends on the sky coordinates $\theta$ and $\varphi$. The only
difference is that $Q = F \cos 2 \chi$ and $U = F \sin 2 \chi$, so that

$$
<Q^2> = <U^2>
$$

\noindent
Here the average is taken over realisation.

\section{The CMB and its Stokes Parameters}

The polarizarion tensor of the CMBR can be obtained as a solution of the
Boltzman equations, which describes the transfer of radiation in
nonstationary plasma and in presence of variable and inhomogeneous
gravitational field  \cite{bas80}, \cite{saz84}, \cite{har93},
\cite{saz95}. The linear polarization of the CMBR is produced mainly at  
the recombination epoch by  Thomson scattering on free electrons in  
primordial cosmological plasma.

The gravitational field in the Universe can be separated into background
gravitational field, that is the homogeneous and isotropic FRW metric, and
inhomogeneous and variable waves of three types: density perturbations,
vector fluctuations, and gravitational waves. In a homogeneous and
isotropic expanding Universe only one parameter of the CMBR is changed:
the temperature, which decreases adiabatically. This expansion does
not produces neither anisotropy nor polarization. Therefore the intensity
$I=I_l + I_r$ decreases adiabatically during the expansion, being this
valid both
for $I_l$ and $I_r$ separately. As a consequence
$Q=0$ and $U = I_u =0$.

On the contrary, the inhomogeneous and variable perturbations of
the gravitational field produce both anisotropy and polarization of the  
CMBR. In this case one can introduce small variations of $I_l$, $I_r$ and
$I_u$ ($\delta_l$, $\delta_r$, $\delta_u$) describing both anisotropy
and polarization of the CMBR.

Here the contribution of only density waves into polarization will be   
considered.

The equations and details of its solution can be found in \cite{saz95},
\cite{saz96a},  \cite{saz96b} and in reference therein.

One can introduce auxiliary functions $\alpha$ and $\beta$:
$$
\begin{array}{c}
\delta_l + \delta_r = (\mu^2 - {1 \over 3}) \alpha, \\
\delta_l - \delta_r = ( 1 - \mu^2) \beta, \\
\delta_u =0,
\end{array}
$$

\noindent
in case of density perturbations taken as single plain wave 
the Boltzman equations can be re-written in terms of these parameters:
$$
\begin{array}{c}
\frac{d \alpha}{d \eta} = F - \frac{9}{10} \sigma_T n_e a(\eta) \alpha -
\frac{6}{10} \sigma_T n_e a(\eta) \beta \\
\frac{d \beta}{d \eta} = - \frac{1}{10} \sigma_T n_e a(\eta) \alpha
- \frac{4}{10} \sigma_T n_e a(\eta) \beta
\end{array}
$$
  
\noindent
Here $F$ is the gravitational force which drives both anisotropy and
polarization, $\sigma_T$ is the Thomson cross-section, $n_e$ is the
density  of free electrons, $a(\eta)$ is the scale factor and $\mu$ is
the angle between the wave vector of the perturbation and the line of
sight.

The solution of these equations, which 
produce  only the Stokes   
parameter $Q$, is:
$$
Q= \frac{1}{7}(1 - \mu^2) \int F(\eta) \left( e^{-\tau} - e^{-{3 \over 10}
\tau}\right)
d\eta
$$

\noindent
where $\tau(\eta)$ is the variable optical depth. In this case the
Stokes parameter $U$ equal to zero.

The distribution of $Q$ over the sky can be obtained by adding
the contribution from all plain waves.

\section{Mathematical Properties of Stokes Parameters}

The Stokes parameters of the CMBR are rank 2 tensor on the sphere. The
rotationally invariant values are $I$ (intensity of radiation), $Q+iU$
and $Q-iU$. The intensity $I$ is decomposed into usual (scalar) spherical
harmonics $Y_{lm}(\theta, \varphi)$.
$$
I=\sum_{l,m} a_{lm} Y_{lm}(\theta, \varphi)
$$

\noindent
Moreover, the two values $Q \pm iU$ can be decomposed into $\pm 2$ spin
harmonics \cite{saz96a}, \cite{saz96b}, \cite{sel97} $ Y_{lm}^{\pm
2}(\theta, \varphi)$:
$$
Q \pm iU=\sum_{l,m} a_{lm}^{\pm 2} Y_{lm}^{\pm 2}(\theta, \varphi)
$$

That form a complete orthonormal system (see, for  instance, \cite{gel58},
\cite{gol67}, \cite{zer70}, \cite{tho80}). They  can
re-written in term of generalized Jacobi polynomials \cite{saz96b},   
\cite{saz99}:
$$
Y^2_{lm}(\theta, \varphi) = N^2_{lm} P^2_{lm}(\theta) e^{i m \varphi}
$$
     
\noindent
$$
Y^{-2}_{lm}(\theta, \varphi) = N^{-2}_{lm} P^{-2}_{lm}(\theta) e^{i m
\varphi}
$$

\noindent
That, as a function of $\varphi$, are similar to scalar spherical 
harmonics. The difference of tensor harmonics lies in $P$, which can be
expressed in terms of Jacobi polynomials:
\begin{equation}
P^s_{lm}(x) = (1 - x)^{\displaystyle (m + s) \over 2} (1 +
x)^{\displaystyle (s - m) \over 2} P^{(m + s, s
- m)}_{lm}(x)
\label{jac}
\end{equation}
  
\noindent
Alternatively, the equivalent polynomials derived in \cite{gol67} can be
used.

In equation (\ref{jac}) $s = \pm 2$ and the normalization factor is:
$$
N^s_{lm} = \frac{1}{2^s}\sqrt{\frac{2l + 1}{4 \pi}}
\sqrt{\frac{(l - s)!(l + s)!}{(l - m)! (l + m)!}}
$$

The harmonics amplitudes $a^{\pm 2}_{lm}$ correspond to the
Fourier spectrum of angular decomposition of
rotationally  invariant combinations of Stokes parameters.

According to \cite{sel97} one can introduce the  $E$ (electric) and $B$
(magnetic) modes of these
harmonics
$$
a^E_{lm} = \frac{1}{2}\left( a^{+2}_{lm} + a^{-2}_{lm}\right)
$$   

\noindent
$$
a^B_{lm} = \frac{i}{2}\left( a^{+2}_{lm} - a^{-2}_{lm}\right)
$$
  
That have different parity. In order to clarify this mathematical
statement let us consider the following situation.

Two observers define the main axes in different way: while the $\vec l$
axes are directed in the same direction, the $\vec r$ axes have different
directions. Let both observers measure the Stokes parameters: the $Q$
parameters look similar to both observers, while $U$  parameters have  
different sign.

It means that by transforming the cordinate system $Oxyz$ into the new
coordinate system $\tilde O \tilde x \tilde y \tilde z$  the vectors
$\vec l$ and $\vec r$ are transformed according to:
$$
\begin{array}{c}
\tilde  {\vec l} = \vec l \\
\tilde  {\vec r} = - \vec r
\end{array}
$$

\noindent
Similarly, the E and B modes are transformed as vectors $\vec
l$ and $\vec r$:
$$
\begin{array}{c}
\tilde a^{E} = a^{E} \\
\tilde a^{B} =- a^{B}
\end{array}
$$

It is necessary to mention that $a^E$ and $a^B$ are not correlated.
  
The density perturbations generate polarization in such a way that $Q \ne
0$ and $U=0$. This is equivalent to electric mode excitation and vanishing
of magnetic modes in the CMBR polarization \cite{sel97}:
\begin{equation}
a^{E}_d \ne 0 \hskip2cm a^{B}_d=0
\end{equation}

Since the synchrotron radiation produces both $Q$ and $U$, 
both electric and magnetic modes exist:
$$
a^{E}_s \ne 0, \hskip2cm a^{B}_s \ne 0
$$

\noindent  
because the synchrotron radiation is connected with the axial vector.
The electric and the magnetic components of
the synchrotron emission obey to the equation:
\begin{equation}
<(a_s^E)^2> = <(a_s^B)^2>
\label{syn}
\end{equation}

\section{Separation of Synchrotron Radiation and the CMB Radiation}

Therefore, to separate the synchrotron polarization from the CMBR
polarization generated by density perturbations one has to separate the
polarized components obeying
$$
a^E \ne 0 \hskip4cm a^B=0
$$

$E$ and $B$ modes of synchrotron foreground do not correlate
each other and they do not correlate with polarization of the CMBR.

To estimate the contribution of $E$ modes generated by density
perturbations one can choose the estimator:
$$
D = <(a^E)^2> - <(a^B)^2>
$$

Let us consider this estimator more detailed: the $a^E$ component is
the sum of two components, $a^E = a^E_s + a^E_d$, and the indexes $s$
and $d$ represent the synchrotron component and the CMBR component,
respectively. The same equation is valid for the $B$ component: $a^B =
a^B_s + a^B_d$. The mean square of the electric component is: $<(a^E)^2> =
<(a^E_s)^2> + <(a^E_d)^2> + 2<a^E_s a^E_d>$. Since  density fluctuations
and synchrotron fluctuations do not correlate the last term vanishes and
the equation can be written as $<(a^E)^2> = <(a^E_s)^2> +
<(a^E_d)^2>$. The same equation is valid for $B$ components.
As far as the squares of both $E$ and $B$ component of the synchrotron
radiation canceled out (\ref{syn}) and the $B$ component of density
fluctuations is equal to zero, the estimator $D$ is equal only to the
square of the $E$ mode of density fluctuations:
$$
D = <(a^E_d)^2>
$$

Thus the $D$ estimator can be used  to separate the polarization of
galactic synchrotron emission from the polarization of the CMBR generated
by density perturbation.

\section*{Acknowledgments}
The author is indebted to Drs. S.Cortiglioni and E.Vinjakin for
many helpfull discussions and suggestions. Also, he would like to
acknowledge the Te.S.R.E. CNR for hospitality during the last stage of
preparation of this paper.

{}

\end{document}